\documentclass[
	journal
]{IEEEtran}

\IEEEoverridecommandlockouts 
\makeatletter
\def\markboth#1#2{\def\leftmark{\@IEEEcompsoconly{\sffamily}\MakeUppercase{\protect#1}}%
\def\rightmark{\@IEEEcompsoconly{\sffamily}\MakeUppercase{\protect#2}}}
\makeatother

\usepackage[amssymb]{SIunits}
\usepackage[cmex10]{amsmath} 
\usepackage{amssymb}
\usepackage{xspace}
\usepackage[latin1]{inputenc}
\usepackage{xcolor}
\usepackage{cite}
\usepackage[pdftex]{graphicx}
\usepackage[caption=false]{subfig}
\usepackage{url}
\usepackage[normalem]{ulem}
\newtheorem{example}{Example}

	\newcommand\xqed[1]{%
  \leavevmode\unskip\penalty9999 \hbox{}\nobreak\hfill
  \quad\hbox{#1}}
\newcommand\demo{\xqed{$\triangle$}}

\newenvironment{DIFnomarkup}{}{}
\graphicspath{%
	{../pstricks/}%
	{../mafig/}%
}%

\usepackage[
	shortcuts,
	nonumberlist,	
	acronym, 
	hyperfirst=true,
	section=section, 
	order=letter,
	numberedsection=nolabel 
]{glossaries}

\newglossary[slg]{los}{syi}{syg}{List of Symbols} 
\makeglossaries 
\glsdisablehyper 
\newglossarystyle{mylist}{
	\glossarystyle{list} 
}

\newglossarystyle{mysymbolstyle}{

	\renewcommand*{\glsgroupheading}[1]{}%
}

\newglossarystyle{myacronymstyle}{
  \glossarystyle{mysymbolstyle} %

}
\newcommand{\NewS}[5][\newcommand]{
	\newglossaryentry{symb:#2}{
		name=\ensuremath{#3},
		description={\nopostdesc #4}, 
		sort=sym#5,
		type=los,
	}
	\glsadd{symb:#2}
	\expandafter\def\csname #2\endcsname{\ensuremath{#3}} 
}

\newcommand{\NewC}[5][\glsadd]{
	\newglossaryentry{symb:#2}{
		name=\ensuremath{#3},
		description={\nopostdesc #4}, 
		sort=sym#5,
		type=los,
	}
	\expandafter\def\csname #2\endcsname{\ensuremath{#3}} 
}

\newcommand{\NewF}[6][\newcommand]{
	\newglossaryentry{symb:#2}{
		name=\ensuremath{#3\cdot#4},
		description={\nopostdesc #5}, 
		sort=sym#6,
		type=los,
	}
	\expandafter\def\csname #2\endcsname ##1{\ensuremath{#3{##1}#4}} 
}
\NewS{TXR}{R_0}{transmitted radius}{rx}
\NewS{txR}{r_0}{transmitted radius}{rx}

\NewS{TXP}{\Theta_0}{transmitted phase}{tx}
\NewS{txP}{\theta_0}{transmitted phase}{tx}

\NewS{RXR}{R}{received radius}{ry}
\NewS{rxR}{r}{received radius}{ry}

\NewS{RXP}{\Theta}{received phase}{ty}
\NewS{rxP}{\theta}{received phase}{ty}

\NewS{TXRdet}{\hat{R}_0}{transmitted radius}{rx}

\NewS{estX}{\hat{X}}{estimated symbol}{xh}
\NewS{esttxR}{\hat{R}_x}{estimated transmitted radius}{rxe}
\NewS{esttxP}{\hat{\Theta}_x}{estimated transmitted phase}{txe}

\NewS{NLPN}{\Phi_{\text{NL}}}{nonlinear phase noise}{phi}
\NewS{Len}{{L}}{total fiber length}{L}
\NewS{Power}{{P}}{input power}{L}

\NewS{X}{X}{transmitted symbol}{x}
\NewS{Y}{Y}{received symbol}{y}
\NewS{setX}{\mathcal{X}}{signal constellation}{xs}
\NewS{setAPSK}{\setX_{\text{APSK}}}{APSK signal set}{xsapsk}

\NewS{rad}{r}{radius}{r}
\NewS{ppr}{l}{points per ring}{l}
\NewS{phaseoffset}{\phi}{phase offset per ring}{p}
\NewS{NumRings}{N}{number of rings in the APSK constelllation}{N}
\NewS{RD}{\bold{r}}{radii distribution}{rr}

\NewS{NumSpans}{K}{number of spans}{k}

\NewS{SEP}{\text{SEP}}{symbol error probability}{ty}

\NewS{labeling}{\mathbb{L}}{labeling}{l}
\NewS{BRGC}{\mathbb{G}}{binary reflected gray code}{gc}
\NewS{ODP}{\otimes}{ordered direct product}{odp}

\NewS{Reg}{\mathcal{R}}{compact bounding region}{reg}

\NewS{imag}{\jmath}{imaginary unit}{j} 

\NewC{Code}{\mathcal{C}}{code}{C:code}
\NewC{vecI}{\boldsymbol{I}}{identity matrix}{i}
\NewC{vecZero}{\boldsymbol{0}}{all-zero vector}{0z}
\NewC{xor}{\oplus}{exclusive or}{0excl}
\NewC{SNR}{\text{SNR}}{signal to noise ratio}{s}
\NewC{mod}{\;\operatorname{mod}\,}{modulo operation}{m}	
\NewC{cconv}{\circledast}{circular convolution}{0}	

\NewC{natural}{\mathbb{N}}{set of natural numbers}{n} 
\NewC{real}{\mathbb{R}}{set of real numbers}{rs}
\NewC{rational}{\mathbb{Q}}{set of rational numbers}{q}
\NewC{integer}{\mathbb{Z}}{set of integer numbers}{zzzz}
\NewC{complex}{\mathbb{C}}{set of complex numbers}{c}

\NewC{Rn}{\mathbb{R}^n}{real Euclidean $n$-dimensional space}{rsn}
\NewC{GF}{\mathbb{F}_2}{Galois field of size two}{f}

\NewC{N}{\mathcal{N}}{normal distribution}{no}
\NewC{Norm}{\mathcal{N}}{normal distribution}{N}
\NewC{Unif}{\operatorname{Unif}}{uniform distribution}{uz}
\NewF{norm}{||}{||}{Euclidean norm}{0e}
\NewF{Qla}{Q_{\Lambda}(}{)}{nearest neighbor lattice quantizer}{quant}
\NewF{Qco}{Q_{\Code}(}{)}{binary quantizer with respect to a linear code $\Code$}{quant2}
\NewF{E}{\mathbb{E}\left[}{\right]}{expectancy operator}{e}
\NewF{bef}{\operatorname{H}(}{)}{binary entropy function}{he}
\NewF{Pr}{\operatorname{Pr}\left[}{\right]}{probability of an event}{pr}

\newcommand{\IE}{i.e., } 
\newcommand{\EG}{e.g., } 

\newcommand{\define}{\triangleq}
\newcommand{\vect}[1]{\boldsymbol{#1}}

\newcommand{\D}{\mathrm{d}}

\renewcommand{\epsilon}{\varepsilon}

\makeatletter
\newcommand{\vast}{\bBigg@{5}}
\makeatother

\definecolor{shadecolor}{rgb}{0.97,0.97,0.97}%
\definecolor{framecolor}{rgb}{0,0,0}%

\newcommand{\abbr}[1]{{#1}}				

\makeatletter
\let\aclOLD=\acl
\renewcommand{\acl}[1]{%
  \begingroup    
  \let\@@underline=\relax
  \aclOLD{#1}%
  \endgroup
}
\makeatother

\newcommand{\NewA}[3]{
	\newacronym{#1}{#2}{#3}
}

\NewA{af}{AF}{\abbr{a}mplify-and-\abbr{f}orward}
\NewA{apsk}{APSK}{\abbr{a}mplitude \abbr{p}hase-\abbr{s}hift \abbr{k}eying}
\NewA{ask}{ASK}{\abbr{a}mplitude-\abbr{s}hift \abbr{k}eying}
\NewA{ase}{ASE}{\abbr{a}mplified \abbr{s}pontaneous \abbr{e}mission}
\NewA{awgn}{AWGN}{\abbr{a}dditive \abbr{w}hite \abbr{G}aussian \abbr{n}oise}
\NewA{biawgn}{BI-AWGN}{binary-input \abbr{a}dditive \abbr{w}hite \abbr{G}aussian \abbr{n}oise}
\NewA{bep}{BEP}{\abbr{b}it \abbr{e}rror \abbr{p}robability}
\NewA{ber}{BER}{\abbr{b}it \abbr{e}rror \abbr{r}ate}
\NewA{qap}{QAP}{\abbr{q}uadratic \abbr{a}assignment \abbr{p}roblem}
\NewA{bicm}{BICM}{\abbr{b}it-\abbr{i}nterleaved \abbr{c}oded \abbr{m}odulation}				
\NewA{cm}{CM}{\abbr{c}oded \abbr{m}odulation}				
\NewA{qpsk}{QPSK}{quadrature phase-shift keying}				
\NewA{mlcm}{MLCM}{multilevel coded modulation}				
\NewA{bpsk}{BPSK}{\abbr{b}inary \abbr{p}hase-\abbr{s}hift \abbr{k}eying}				
\NewA{bsc}{BSC}{\abbr{b}inary \abbr{s}ymmetric \abbr{c}hannel}				
\NewA{brgc}{BRGC}{\abbr{b}inary \abbr{r}eflected \abbr{G}ray \abbr{c}ode}				
\NewA{cf}{CF}{\abbr{c}haracteristic \abbr{f}unction}
\NewA{csit}{CSIT}{\abbr{c}annnel \abbr{s}tate \abbr{i}nformation at the \abbr{transmitter}}
\NewA{csi}{CSI}{\abbr{c}annnel \abbr{s}tate \abbr{i}nformation}
\NewA{df}{DF}{\abbr{d}ecode-and-\abbr{f}orward}
\NewA{fd}{FD}{\abbr{f}ull-\abbr{d}uplex}
\NewA{fft}{FFT}{\abbr{f}ast \abbr{F}ourier \abbr{t}ransform}
\NewA{iid}{IID}{independent and \abbr{i}dentically \abbr{d}istributed}
\NewA{isi}{ISI}{\abbr{i}nter\abbr{s}ymbol \abbr{i}nterference}
\NewA{lb}{LB}{\abbr{l}ower \abbr{b}ound}
\NewA{map}{MAP}{\abbr{m}aximum \abbr{a} \abbr{p}osteriori}
\NewA{mf}{MF}{\abbr{m}odulo-and-\abbr{f}orward}
\NewA{mlan}{MLAN}{\abbr{m}odulo-\abbr{l}attice \abbr{a}dditive \abbr{n}oise}
\NewA{ml}{ML}{\abbr{m}aximum \abbr{l}ikelihood}
\NewA{mmse}{MMSE}{\abbr{m}inimum \abbr{m}ean \abbr{s}quare \abbr{e}rror}
\NewA{nlpc}{NLPC}{\abbr{n}onlinear \abbr{p}hase \abbr{c}ompensation}
\NewA{nlpn}{NLPN}{\abbr{n}onlinear \abbr{p}hase \abbr{n}oise}
\NewA{nc}{NC}{\abbr{n}etwork \abbr{c}oding}
\NewA{ofmd}{OFDM}{\abbr{o}rthogonal \abbr{f}requency-\abbr{d}ivision \abbr{m}ultiplexing}			
\NewA{dp}{DP}{\abbr{d}ual-\abbr{p}olarization}
\NewA{pam}{PAM}{\abbr{p}ulse \abbr{a}mplitude \abbr{m}odulation}
\NewA{pdf}{PDF}{\abbr{p}robability \abbr{d}ensity \abbr{f}unction}
\NewA{plnc}{PNC}{\abbr{p}hysical-layer \abbr{n}etwork \abbr{c}oding}
\NewA{psk}{PSK}{\abbr{p}hase-\abbr{s}hift \abbr{k}eying}
\NewA{pmd}{PMD}{\abbr{p}olarization \abbr{m}ode \abbr{d}ispersion}
\NewA{pm}{PM}{\abbr{p}olarization-\abbr{m}ultiplexed}
\NewA{pdm}{PDM}{\abbr{p}olarization-\abbr{d}ivision \abbr{m}ultiplexing}
\NewA{qam}{QAM}{\abbr{q}uadrature \abbr{a}mplitude \abbr{m}odulation}
\NewA{sqp}{SQP}{\abbr{s}equential \abbr{q}uadratic \abbr{p}rogramming}
\NewA{rd}{RD}{\abbr{r}adii \abbr{d}istribution}
\NewA{sep}{SEP}{\abbr{s}ymbol \abbr{e}rror \abbr{p}robability}
\NewA{ser}{SER}{\abbr{s}ymbol \abbr{e}rror \abbr{r}ate}
\NewA{si}{SI}{\abbr{s}ide \abbr{i}nformation}
\NewA{sp}{SP}{\abbr{s}ingle-\abbr{p}olarization}
\NewA{sanr}{SNR}{\abbr{s}ignal-to-(additive-)\abbr{n}oise \abbr{r}atio}
\NewA{snr}{SNR}{\abbr{s}ignal-to-\abbr{n}oise \abbr{r}atio}
\NewA{snlse}{sNLSE}{\abbr{s}tochastic \abbr{n}onlinear \abbr{S}chr\"odinger \abbr{e}quation}
\NewA{nlse}{NLSE}{\abbr{n}onlinear \abbr{S}chr\"odinger \abbr{e}quation}
\NewA{stwrc}{sTRC}{\abbr{s}eparated \abbr{t}wo-way \abbr{r}elay \abbr{c}hannel}
\NewA{stwtrc}{sTTRC}{\abbr{s}eparated \abbr{t}wo-way \abbr{t}wo-\abbr{r}elay \abbr{c}hannel}
\NewA{ts}{TS}{\abbr{t}wo-\abbr{s}tage}
\NewA{twrc}{TRC}{\abbr{t}wo-way \abbr{r}elay \abbr{c}hannel}
\NewA{twtrc}{TTRC}{\abbr{t}wo-way \abbr{t}wo-\abbr{r}elay \abbr{c}hannel}
\NewA{wdm}{WDM}{\abbr{w}avelength-\abbr{d}ivision \abbr{m}ultiplexing}
\NewA{vn}{VN}{variable node}
\NewA{cn}{CN}{constraint node}
\NewA{de}{DE}{density evolution}
\NewA{sc}{SC}{spatially-coupled}
\NewA{ldpc}{LDPC}{low-density parity-check}
\NewA{bp}{BP}{belief propagation}
\NewA{dm}{DM}{dispersion-managed}
\NewA{spm}{SPM}{self-phase modulation}
\NewA{xpm}{XPM}{cross-phase modulation}
\NewA{fwm}{FWM}{four-wave-mixing}
\NewA{ixpm}{IXPM}{intrachannel cross-phase modulation}
\NewA{ifwm}{IFWM}{intrachannel four-wave-mixing}
\NewA{ssfm}{SSFM}{split-step Fourier method}
\NewA{fec}{FEC}{forward error correction}
\NewA{psd}{PSD}{power spectral density}
\NewA{ook}{OOK}{on-off keying}
\NewA{smf}{SMF}{single-mode fiber}
\NewA{ssmf}{SSMF}{standard single-mode fiber}
\NewA{dbp}{DBP}{digital backpropagation}
\NewA{edfa}{EDFA}{erbium-doped fiber amplifier}
\NewA{ofdm}{OFDM}{orthogonal frequency division multiplexing}
\NewA{exit}{EXIT}{extrinsic information transfer}
\NewA{pexit}{P-EXIT}{protograph extrinsic information transfer}
\NewA{osnr}{OSNR}{optical signal-to-noise ratio}
\NewA{roadm}{ROADM}{reconfigurable optical add-drop multiplexer}
\NewA{rps}{RPS}{Raman pump station}
\NewA{mlse}{MLSE}{maximum likelihood sequence estimation}
\NewA{dcf}{DCF}{dispersion compensating fiber}
\NewA{tcm}{TCM}{trellis coded modulation}
\NewA{edc}{EDC}{electronic dispersion compensation}
\NewA{ldpcc}{LDPCC}{low-density parity-check convolutional}
\NewA{llr}{LLR}{log-likelihood ratio}
\NewA{ra}{RA}{repeat-accumulate}
\NewA{ira}{IRA}{irregular-repeat-accumulate}
\NewA{ara}{ARA}{accumulate-repeat-accumulate}
\NewA{mi}{MI}{mutual information}
\NewA{vdmm}{VDMM}{variable degree matched mapping}
\NewA{gmi}{GMI}{generalized mutual information}
\NewA{wd}{WD}{windowed decoder}
\NewA{gn}{GN}{Gaussian noise}

\NewA{hd}{HD}{hard-decision}
\NewA{hdd}{HDD}{hard-decision decoding}
\NewA{sd}{SD}{soft-decision}
\NewA{sdd}{SDD}{soft-decision decoding}

\NewA{emp}{EMP}{extrinsic message passing}
\NewA{imp}{IMP}{intrinsic message passing}

\NewA{gldpc}{GLDPC}{generalized low-density parity-check}
\NewA{scgldpc}{SC-GLDPC}{spatially-coupled generalized low-density parity-check}

\NewA{scldpc}{SC-LDPC}{spatially-coupled low-density parity-check}
\NewA{scfec}{SC-FEC}{spatially-coupled forward error correction}

\NewA{tbbc}{TBBC}{tightly-braided block code}

\NewA{gpc}{GPC}{generalized product code}
\NewA{otn}{OTN}{optical transport network}
\NewA{bch}{BCH}{Bose--Chaudhuri--Hocquenghem}
\NewA{bbbc}{BBBC}{block-wise braided block code}

\NewA{hpc}{HPC}{``half'' product code}
\NewA{pc}{PC}{product code}
\NewA{rv}{RV}{random variable}

\NewA{oh}{OH}{overhead}

\NewA{bdd}{BDD}{bounded-distance decoding}
\NewA{ncg}{NCG}{net coding gain}
\NewA{gwbp}{Galton--Watson branching process}{Galton--Watson branching process}

\newacronym[%
	longplural={binary erasure channels},%
	shortplural={BECs}%
]{bec}{BEC}{binary erasure channel}%

\renewcommand{\imag}{\ensuremath{j}}
\renewcommand{\vect}[1]{\ensuremath{\mathbf{#1}}}

\newcommand{\VNs}{\glspl{vn}\xspace}
\newcommand{\CNs}{\glspl{cn}\xspace}
\newcommand{\CnNum}{r_{\mathcal{C}}}

\newcommand{\lmax}{l_\text{max}}

\newcommand{\eps}{\varepsilon}

\newcommand{\conv}{*}

\newcommand{\Lsp}{\ensuremath{L_{\text{sp}}}}
\newcommand{\Nsp}{\ensuremath{N_\text{sp}}}
\newcommand{\SEF}{\ensuremath{n_\text{sp}}}

\newcommand{\PSDl}{\ensuremath{{N}_{\text{EDFA}}}}

\newcommand{\PolX}{\ensuremath{\mathsf{x}}}
\newcommand{\PolY}{\ensuremath{\mathsf{y}}}

\newcommand{\LF}{\ensuremath{{M}}} 
\newcommand{\TL}{\ensuremath{{T}}} 
\newcommand{\WS}{\ensuremath{{W}}} 

\newcommand{\maxIter}{\ensuremath{l_\text{max}}}

\newcommand{\proto}{\ensuremath{\mathbf{P}}}

\newcommand{\Rldpc}{\ensuremath{R}}
\newcommand{\Rgldpc}{\ensuremath{R'}}

\newcommand{\kldpc}{\ensuremath{k_{\mathcal{C}}}}
\newcommand{\nldpc}{\ensuremath{n_{\mathcal{C}}}}

\newcommand{\Cbch}{\ensuremath{\mathcal{B}}}
\newcommand{\nbch}{\ensuremath{n_{\Cbch}}}
\newcommand{\kbch}{\ensuremath{k_{\Cbch}}}

\newcommand{\CnNumSp}{\ensuremath{{m}_{\text{c}}}} 

\newcommand{\RevCR}[1]{}

\begin{document}

\begin{DIFnomarkup}



\title{Terminated and Tailbiting Spatially-Coupled Codes with
Optimized Bit Mappings for Spectrally Efficient Fiber-Optical Systems}


\author{
	\IEEEauthorblockN{
	Christian Häger, \emph{Student Member, IEEE},
	Alexandre Graell i Amat, \emph{Senior Member, IEEE},
	Fredrik Brännström, \emph{Member, IEEE},
	Alex Alvarado, \emph{Member, IEEE}, and
	Erik Agrell, \emph{Senior Member, IEEE}
	\thanks{Parts of this paper have been presented at the
	European Conference on Optical Communication (ECOC), Cannes, France,
	Sep. 2014.}
	\thanks{This work was partially funded by the Swedish Research Council under
	grant \#2011-5961 and by the 
  Engineering and Physical Sciences Research Council (EPSRC) project UNLOC (EP/J017582/1), United Kingdom. The
	simulations were performed in part on resources provided by the
	Swedish National Infrastructure for Computing (SNIC) at C3SE. } 
	\thanks{C.~Häger, A.~Graell i Amat, F.~Brännström, and E.~Agrell are with the
	Department of Signals and Systems, Chalmers University of Technology, Gothenburg,
	Sweden (emails: \{christian.haeger, alexandre.graell,
	fredrik.brannstrom,
	agrell\}@chalmers.se). }
	\thanks{A.~Alvarado is with the Optical Networks
	Group, Department of Electronic and Electrical Engineering,
	University College London, London WC1E7JE, UK (email:
	alex.alvarado@ieee.org).}
	}
}

\maketitle

\end{DIFnomarkup}

\begin{abstract}
	We study the design of spectrally efficient fiber-optical
	communication systems based on different spatially-coupled (SC)
	forward error correction (FEC) schemes. In particular, we optimize
	the allocation of the coded bits from the FEC encoder to the
	modulation bits of the signal constellation. Two SC code classes are
	considered. The codes in the first class are protograph-based
	low-density parity-check (LDPC) codes which are decoded using
	iterative soft-decision decoding. The codes in the second class are
	generalized LDPC codes which are decoded using iterative
	hard-decision decoding. For both code classes, the bit allocation is
	optimized for the terminated and tailbiting SC cases based on a
	density evolution analysis. An optimized bit allocation can
	significantly improve the performance of tailbiting SC codes codes
	over the baseline sequential allocation, up to the point where they
	have a comparable gap to capacity as their terminated counterparts,
	at a lower FEC overhead. For the considered terminated SC codes, the
	optimization only results in marginal performance improvements,
	suggesting that in this case a sequential allocation is close to
	optimal. 
\end{abstract}



\glsresetall 

%
%

\section{Introduction}
\label{sec:introduction}

\begin{figure*}[t]
	\centering
		\includegraphics{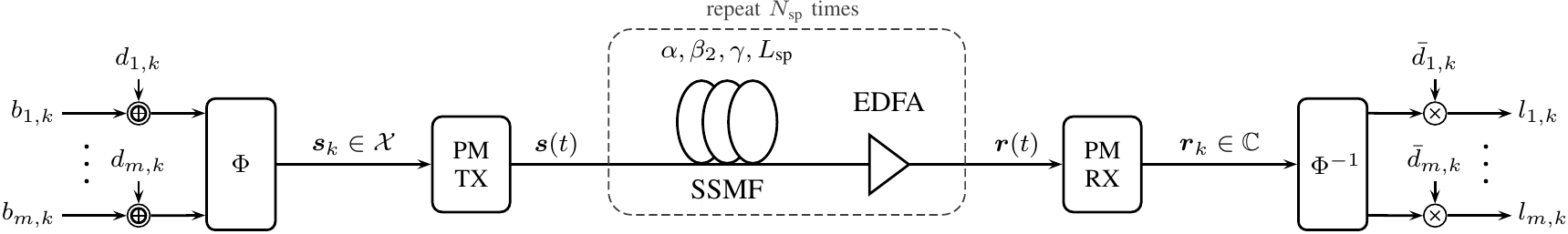}
	 \caption{Block diagram of the considered PM transmission system. }
	 \label{fig:BlockDiagramConf}
\end{figure*}

Designing spectrally efficient fiber-optical systems that can operate
close to the capacity limits \cite{Essiambre2010} has become an
important research topic \cite{Smith2010, Schmalen2013, Beygi2014}.
Such systems are often implemented according to the pragmatic
\gls{bicm} paradigm \cite{Caire1998}, where a single binary \gls{fec}
encoder is used in combination with a nonbinary signal constellation.
A random allocation (or interleaving) \cite{Caire1998} of
the coded bits from the FEC encoder to the modulation bits of the
signal constellation is commonly assumed. In this paper, we optimize
the allocation to the modulation bits for a coherent long-haul
\gls{pm} fiber-optical system. In particular, we consider different
\gls{sc} \gls{fec} schemes both with \gls{sdd} and \gls{hdd}. 


SC low-density parity-check (SC-LDPC)\glsunset{ldpc}\glsunset{scldpc}
codes have attracted a great deal of attention in the recent years.
They are considered as viable candidates for future spectrally
efficient fiber-optical systems \cite{Schmalen2013, Leven2013,
Miyata2013} due to their capacity-achieving performance for many
communication channels \cite{Kudekar2011}. \gls{scldpc} codes promise
excellent \gls{bp} performance with a quasi-regular node degree
distribution and low node degrees. The \gls{bp} performance of
\gls{scldpc} codes can further be improved by increasing the node
degrees, whereas the decoding performance for regular \gls{ldpc} codes
generally worsens if the node degrees are increased
\cite{Kudekar2011}. While irregular \gls{ldpc} codes can also perform
close to capacity \cite{Sae-YoungChung2001}, the optimal degree
distribution depends on the code rate and/or the channel
\cite{Richardson2001a}. High node degrees are also often required for
good performance which leads to a high decoding complexity.





We consider two different \gls{sc} code classes taken from the
literature. The codes in the first class are protograph-based
\gls{scldpc} codes \cite{Thorpe2005,Mitchell2011} which are
decoded using iterative \gls{sdd} in the form of \gls{bp} decoding.
\gls{bp} is a message passing algorithm in which ``soft'' (i.e.,
real-valued) messages are exchanged between the \VNs and \CNs in the
Tanner graph representing the code. The codes in the second class are
\gls{sc} generalized \gls{ldpc} (SC-GLDPC)
\glsunset{scgldpc}\glsunset{gldpc} codes where each coded bit is
protected by two $t$-error correcting \gls{bch} component
codes \cite{Jian2012}. These codes are decoded using iterative
\gls{hdd} with \gls{bdd} of the component \gls{bch} codes.  Iterative
\gls{hdd} can be seen as a message passing algorithm with ``hard''
(i.e., binary) messages in the Tanner graph representing the
\gls{gldpc} code and is significantly less complex than \gls{sdd}
\cite{Smith2012a}.   




The adoption of \gls{sdd} is considered one of the most important
factors for increasing the performance of fiber-optical systems
\cite{Leven2014a}. However, \gls{sdd} poses implementation challenges
at very high data rates motivating the use of less complex \gls{fec}
schemes \cite{Smith2012a}. The \gls{scgldpc} codes we consider in this
paper were proposed in \cite{Jian2012}, where it is shown that they
can approach the capacity of the \gls{bsc} under iterative \gls{hdd}
for high code rates (i.e., low \gls{fec} \glspl{oh}). We use these
codes because a \gls{de} analysis is readily available in
\cite{Jian2012}. This allows us to apply the optimization techniques
for protograph-based \gls{scldpc} codes we previously presented in
\cite{Hager2014a} to the practically relevant case of \gls{scgldpc}
codes with iterative \gls{hdd}. The \gls{scgldpc} code ensemble in
\cite{Jian2012} is closely related to other recently proposed
\gls{fec} schemes for optical transport networks, such as staircase
codes \cite{Smith2012a} (which are themselves related to
\acrlongpl{bbbc} \cite{Lentmaier}), and the modified construction of
\acrlongpl{tbbc} proposed in \cite{Jian2014}. For other related works
on \gls{gldpc} codes for fiber-optical communications, we refer the
interested reader to \cite{Djordjevic2005b, Djordjevic2008} and
references therein.




The outstanding performance of \gls{sc} codes is due to a termination
boundary effect which initiates a wave-like decoding behavior
\cite{Kudekar2011}. This behavior of terminated \gls{sc} codes comes
at the price of a rate loss, \IE a larger FEC \gls{oh}, compared to
the underlying uncoupled codes. So-called tailbiting \gls{sc} codes
provide an interesting solution to this problem, since they do not
suffer from an increased \gls{oh}. However, by default, a tailbiting
\gls{sc} code behaves essentially the same as the underlying uncoupled
code due to the absence of a termination boundary. The main aim of
this paper is to demonstrate that the unequal error protection offered
by the modulation bits of a nonbinary signal constellation can be
exploited to create an artificial termination boundary. This
significantly improves the performance of tailbiting \gls{sc} codes,
both in the case of \gls{sdd} and \gls{hdd}. With an optimized bit
allocation, the capacity gap of the considered tailbiting
\gls{sc} codes is comparable to the gap of their terminated
counterparts, at a lower \gls{fec} \gls{oh}. For the considered
terminated \gls{sc} codes, the performance gain due to an optimized
bit allocation is limited, in particular for the \gls{scgldpc} codes
with \gls{hdd}. Simulation results for both linear and nonlinear
transmission scenarios confirm the \gls{de} analysis. 

The remainder of the paper is organized as follows. In Section
\ref{sec:system_model}, the assumed \gls{pm} transmission system is
described. The two \gls{sc} \gls{fec} schemes are covered in Sections
\ref{sec:scldpc} and \ref{sec:scgldpc}, where we explain the code
construction, the decoding algorithms, and the \gls{de} analysis with
the help of several examples. In Section \ref{sec:optimization}, we
briefly review the optimization techniques for the bit
allocation described in \cite{Hager2014a}, which apply to the
considered \gls{scldpc} codes with \gls{sdd}. We also discuss how they
are easily extended to the \gls{scgldpc} codes with \gls{hdd}.
Results are presented and discussed in Section \ref{sec:results} and
the paper is concluded in Section \ref{sec:conclusion}.








\section{System Model}
\label{sec:system_model}

A block diagram of the considered \gls{pm} fiber-optical transmission
system is shown in Fig.~\ref{fig:BlockDiagramConf}. At each discrete
time instant $k$, the modulator $\Phi$ takes $m$ bits $b_{i,k}$, $i=1,
\ldots, m$, and maps them to a symbol $\boldsymbol{s}_k = (s_{\PolX,
k}, s_{\PolY, k})$ taken with uniform probabilities from a signal
constellation $\mathcal{X} \subset \mathbb{C}^2$ ($|\mathcal{X}| =
2^m$) according to the binary labeling. \RevCR{As an example, the real part
in one polarization corresponding to Gray-labeled PM-64-QAM is shown
in Fig.~\ref{fig:8pam_brgcConf}, where $m=12$.} The modulo-2 addition
of the independent and uniformly distributed bits $d_{i,k}$ (and the
multiplication by $\bar{d}_{i,k} = (-1)^{d_{i,k}}$ at the receiver)
shown in Fig.~\ref{fig:BlockDiagramConf} serves as a symmetrization
technique \cite{Hou2003}.\footnote{The symmetrization makes the bit
error probability independent of the transmitted bits.  This is an
important requirement for the all-zero codeword assumption which is
commonly made in \gls{de} \cite[p.~389]{Ryan2009}.}

The baseband signal in polarization $\PolX$ is $s_{\PolX}(t) = \sum_k
s_{\PolX, k} p(t-k/R_s)$ with (real-valued) pulse shape $p(t)$ and
symbol rate $R_s$ (and similarly for polarization $\PolY$). The
transmit power $P = \lim_{T \to \infty} \int_{-T}^{T} s_{\PolX}(t)^2
\D t/(2T)$ is assumed to be equal in both polarizations. The \gls{pm}
signal $\boldsymbol{s}(t) = (s_\PolX(t), s_\PolY(t))$ is launched into
the fiber and propagates according to the Manakov equation
\cite{Agrawal2005}. The optical link consists of $\Nsp$ spans of
\gls{ssmf} with attenuation coefficient $\alpha$, group velocity
dispersion $\beta_2$, nonlinear Kerr parameter $\gamma$, length
$\Lsp$, and a lumped amplification scheme (no optical dispersion
compensation is assumed). Each \gls{edfa} introduces circularly
symmetric complex Gaussian noise with two-sided power spectral density
$\PSDl = (e^{\alpha \Lsp} - 1) h \nu_s \SEF$ per
polarization\cite{Essiambre2010}, where $h$ is Planck's constant,
$\nu_\text{s}$ the carrier frequency, and $\SEF$ the spontaneous
emission factor. A coherent linear receiver according to $r_{{\PolX},
k} = \left. r_{\PolX}(t) \conv h(t) \conv p(-t) \right|_{t = k / R_s}$
is used in each polarization, where $\conv$ denotes convolution and
$h(t)$ is the impulse response of an equalizer which accounts for
linear distortions due to chromatic dispersion. The frequency response
of the equalizer is given by $H(f) = \exp(\imag 2 \beta_2 \pi^2 f^2
\Nsp \Lsp)$. The received symbles are denoted by $\boldsymbol{r}_k =
(r_{\PolX, k}, r_{\PolY, k})$. 

Two different demodulators $\Phi^{-1}$ are considered. For \gls{sdd},
the demodulator computes ``soft'' reliability information about the
bits $b_{i,k}$ in the form of \glspl{llr} $l_{i,k}$. For \gls{hdd},
the demodulator performs a minimum distance symbol-by-symbol detection
of the received symbols with respect to the signal constellation
$\mathcal{X}$ and outputs the binary labeling associated with the
detected symbol. Both demodulators are based on the assumption that
the discrete-time channel from $\boldsymbol{s}_k$ to
$\boldsymbol{r}_k$ is the \gls{awgn} channel with \gls{snr} denoted by
$\rho$. This assumption is accurate for linear transmission
(i.e., $\gamma = 0$) where $\rho = P/(\Nsp \PSDl R_s)$. For the
considered setup without optical inline dispersion compensation, it
has been shown that this assumption is also justified,
provided that dispersive effects are dominant and nonlinear effects
are weak \cite{Beygi2012, Carena2014}. For this case, $\rho$ can be
computed using \cite[eq.~(15)]{Beygi2012} assuming single channel
transmission. Under the Gaussian Noise model assumption, see
\cite{Carena2014} and references therein, similar expressions for the
SNR are also computable for \acl{wdm} systems. 

\begin{figure}[tb]
	\centering
		\includegraphics{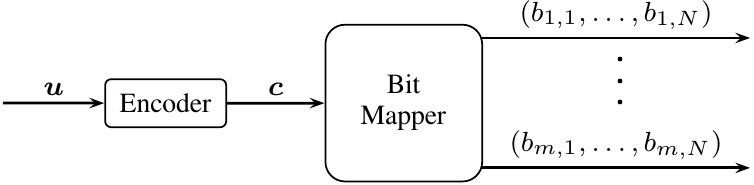}
	\caption{Illustration of the Bit Mapper.}
	\label{fig:bit_mapperConf}
\end{figure}

We consider a system according to the \gls{bicm} paradigm, where a
binary code $\mathcal{C} \subset \{0,1\}^{\nldpc}$ of length $\nldpc$
and dimension $\kldpc$ is employed and each codeword $\boldsymbol{c} =
(c_1, \dots, c_{\nldpc})$ is transmitted using $N = \nldpc/m$ symbols
$\boldsymbol{s}_k$. The allocation of the coded bits to the modulation
bits is determined by a bit mapper\footnote{The bit mapper
should not be confused with the modulator $\Phi$, which is sometimes
also referred to as a mapper. In the literature, the term ``bit
interleaver'' is also frequently used instead of ``bit mapper''.} as
shown in Fig.~\ref{fig:bit_mapperConf}, where $\boldsymbol{u} = (u_1,
\dots, u_{\kldpc})$ is the information word.  The bit mapper
optimization is discussed in Section \ref{sec:optimization}. The
optimization is based on the AWGN channel model because a direct
optimization using \gls{de} for the optical channel defined by the
nonlinear Schr\"odinger equation is not feasible. The accuracy of this
approach is verified through simulation results for a nonlinear
transmission scenario in Section \ref{sec:results_scldpc}. To the best
of our knowledge, there are no comparable works by other authors on
bit mapper optimization for \gls{sc} codes. Hence, as a baseline for a
comparison, we use a sequential mapper according to $b_{i,k} =
c_{(k-1)m+i}$ for $1\leq i \leq m$, $1\leq k \leq N$. For the
considered codes, a sequential mapper has the same expected
performance as a random mapper. 



\section{Protograph-Based SC-LDPC Codes}
\label{sec:scldpc}

\subsection{Code Construction}
\label{sec:scldpc_construction}

An \gls{ldpc} code of length $\nldpc$ and dimension $\kldpc$ is
defined via a sparse parity-check matrix $\bold{H} = [h_{i,j}] \in
\{0,1\}^{\CnNum \times \nldpc}$, where $\CnNum \geq \nldpc-\kldpc$
with equality if and only if $\bold{H}$ has full rank.  \RevCR{The
code can be represented by using a bipartite Tanner graph consisting
of $\nldpc$ \VNs and $\CnNum$ \CNs, where the $i$th \gls{cn} is
connected to the $j$th \gls{vn} if $h_{i,j} = 1$.} One popular method
to construct \gls{ldpc} codes is by using protographs
\cite{Thorpe2005}. A protograph is a bipartite graph defined by an
adjacency matrix $\proto = [p_{i,j}] \in \mathbb{N}_0^{\CnNum' \times
\nldpc'}$, called the base matrix, where $\mathbb{N}_0$ is the set of
nonnegative integers. Given $\proto$, the parity-check matrix
$\vect{H}$ is obtained by replacing each entry $p_{i,j}$ in $\proto$
with a random binary $\LF$-by-$\LF$ matrix which contains $p_{i,j}$
ones in each row and column. This procedure is called lifting and $\LF
\geq \max_{i,j} p_{i,j}$ is the so-called lifting factor.
\RevCR{Graphically, this construction amounts to copying the
protograph $\LF$ times and subsequently permuting edges.  Parallel
edges, \IE $p_{i,j}>1$, are permitted in the protograph and are
resolved in the lifting procedure. } The design rate of the code is
given by $R = 1 - \CnNum/\nldpc = 1 - \CnNum'/\nldpc'$, where $\CnNum
= \CnNum' \LF$ and $\nldpc = \nldpc' \LF$. 

\begin{figure}[tb]
	\centering
		\subfloat[terminated]{\includegraphics{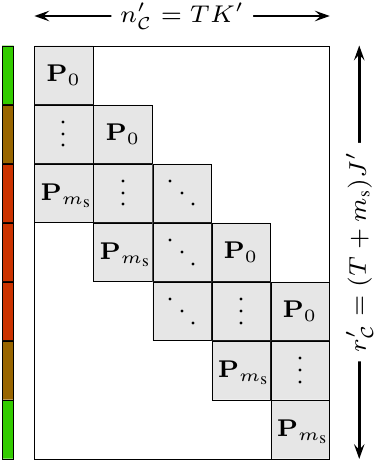}}
		\qquad
		\subfloat[tailbiting]{\includegraphics{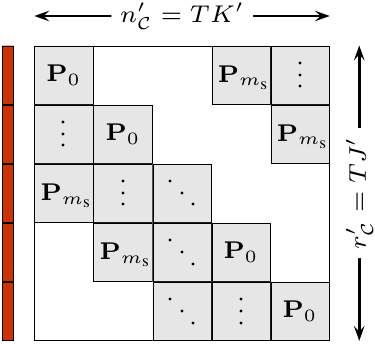}}
	\caption{Base matrices $\proto$ for protograph-based SC-LDPC codes.}
	\label{fig:parity_check}
\end{figure}

\gls{scldpc} codes have parity-check matrices with a band-diagonal
structure and can be constructed using protographs
\cite{Mitchell2011}. The base matrix of a $(J,K)$ regular \gls{scldpc}
code with spatial length $\TL$ is constructed by specifying matrices
$\proto_i$, $0 \leq i \leq m_\text{s}$, of dimension $J'$ by $K'$,
where $m_\text{s}$ is referred to as the memory. The matrices are such
that $\sum_{i=0}^{m_\text{s}} \proto_i$ has column weight $J$ and row
weight $K$ for all columns and rows. \RevCR{(The matrix
$\sum_{i=0}^{m_\text{s}} \proto_i$ can thus be seen as the base matrix
of a regular \gls{ldpc} code with \gls{cn} degree $J$ and \gls{vn}
degree $K$.)} Given the matrices $\proto_i$ and the spatial length
$\TL$, one can construct $\proto$ as shown in
Fig.~\ref{fig:parity_check} (a) for the terminated case and in
Fig.~\ref{fig:parity_check} (b) for the tailbiting case.\footnote{The
terminology originates from the trellis representation of
convolutional codes, where the initial and final states are either
determined by known bits (terminated) or forced to be identical
(tailbiting).} Terminated and tailbiting \gls{scldpc} codes have
design rates $\Rldpc(\TL) = 1 - J'/K' - m_\text{s} J' / (\TL K')$ and
$\Rldpc = 1 - J'/K'$, respectively \cite{Mitchell2011}. The rate loss
for the terminated code with respect to the tailbiting (or the
underlying uncoupled regular) code can be made arbitrary small by
letting $\TL \to \infty$, but this also leads to very long block
lengths $\nldpc = \TL K' \LF$ (assuming a fixed lifting factor $\LF$). 

\begin{figure}[tb]
	\centering
		\subfloat[terminated]{\includegraphics{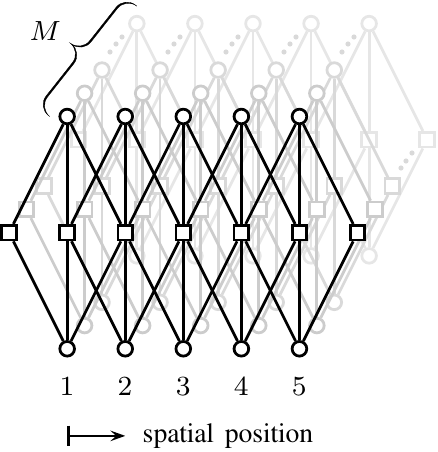}}
		\qquad
		\subfloat[tailbiting]{\includegraphics{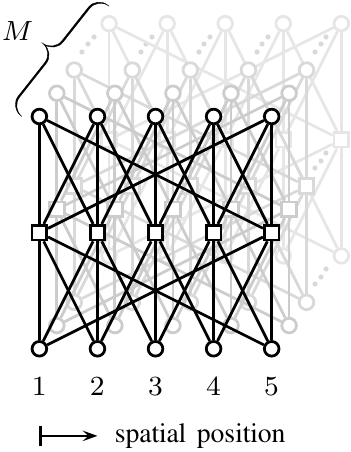}}

		\caption{Protographs for the \gls{scldpc} code with $\TL = 5$ in
		Example \ref{ex:regldpc}. The first step of the lifting procedure
		to obtain the Tanner graph (i.e., copying the protograph $M$
		times) is indicated in light gray.  }
	\label{fig:protograph_lifting}
\end{figure}

\begin{example}
	\label{ex:regldpc}
	Consider the $(3,6)$ regular \gls{scldpc} code with $\proto_0 =
	\proto_1 = \proto_2 = (1,1)$, $\TL = 5$, 
	$J'=1$, $K'=2$, and $m_\text{s} = 2$. The two protographs
	corresponding to the terminated and tailbiting cases are shown in
	Fig.~\ref{fig:protograph_lifting}. The design rates are 
	$\Rldpc(5) = 0.3$ and $\Rldpc = 0.5$, respectively. \demo
\end{example}

\subsection{Soft-Decision Decoding and Density Evolution}
\label{sec:scldpc_decoding}

The protograph-based \gls{scldpc} codes are decoded using the standard
\gls{bp} decoding \cite[Sec.~5.4]{Ryan2009}. In order to alleviate the
long decoding delay and high decoding complexity of \gls{scldpc} codes
under full \gls{bp} decoding, we employ the \gls{wd} with a
window size $\WS$ developed in \cite{Iyengar2012}. \RevCR{The
\gls{wd} restricts message updates to a subset of \VNs and \CNs in the
entire graph depending on the window size $\WS < \TL$. After
$\maxIter$ decoding iterations, this subset changes and the decoding
window slides to the next position.} The \gls{wd} reduces the decoding
delay for terminated \gls{scldpc} codes from $\TL \LF K'$ to
$\WS \LF K'$ coded bits \cite{Iyengar2012}. For tailbiting
\gls{scldpc} codes, additional memory for $(m_\text{s} + W - 1) \LF
K'$ values is required compared to terminated \gls{scldpc} codes, in
order to take the circular wrap-around of the parity-check matrix into
account. In particular, assume that the decoding starts when the
channel observations corresponding to spatial positions $1$ to $m_s +
W$ are received and the first targeted symbols are at position $m_s +
1$. (Due to the circular structure, the last targeted symbols are at
position $m_s$.) Then, the observations corresponding to the first
$m_s$ positions as well as the final \glspl{llr} for the bits at
positions $m_s + 1$ to $m_s + W - 1$ have to be stored. We also point
the interested reader to \cite{Tavares2007a}, where the decoding of
tailbiting \gls{scldpc} codes based on a pipeline decoder architecture
is discussed.



The main tool for the analysis of \gls{ldpc} codes under \gls{bp}
decoding is \gls{de} \cite{Richardson2001}. \gls{de} mimics the
decoding process under a cycle-free graph assumption by tracking how
the densities of the \glspl{llr} evolve with iterations. Tracking the
full densities (or quantized densities in practice) is computationally
demanding and \gls{exit} functions \cite{TenBrink1999} are usually
considered to be a good compromise between computational efficiency
and accuracy. For the protograph-based codes, we employ a modified
protograph EXIT (P-EXIT)\glsunset{pexit} analysis \cite{Liva2007}
which accounts for the different protection levels of a nonbinary
signal constellation and the \gls{wd}, see \cite[Algorithm
1]{Hager2014a}.



\begin{figure}[tb]
	\centering 
	\subfloat[]{\includegraphics{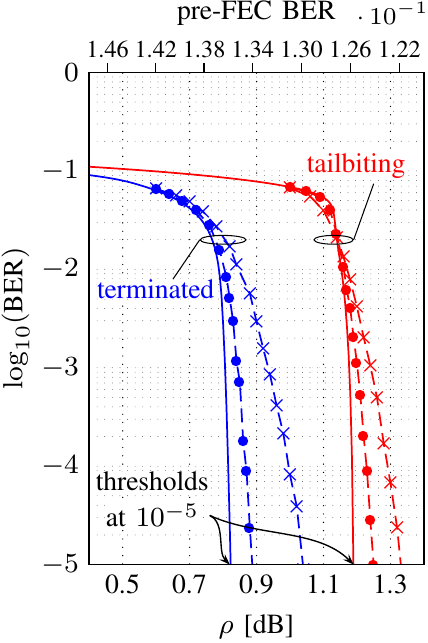}} 
	\quad
	\subfloat[]{\includegraphics{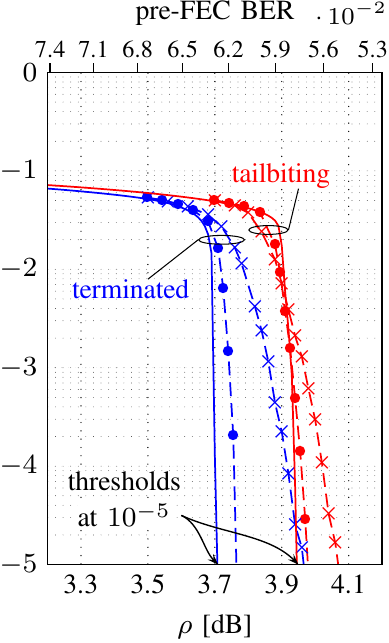}}
	\caption{Predicted (solid lines) and finite-length (dashed lines)
	performance for the codes in Examples \ref{ex:exitldpc} (left) and
	\ref{ex:gldpc} (right). The codes have lengths $80\,000$ (a,
	crosses), $400\,000$ (a, dots), $168\,000$
	(b, crosses), and $1\,260\,000$ (b, dots).  }
	\label{fig:ProtographExample}
\end{figure}

\begin{example}
	\label{ex:exitldpc}
	Consider the $(3,6)$ regular \gls{scldpc} code with $\proto_0 = (2,
	2)$, $\proto_1 = (1,1)$, $\TL = 20$, $J'=1$, $K'=2$, and $m_\text{s}
	= 1$, with rates  $\Rldpc(\TL) = 0.475$ and $\Rldpc = 0.5$,
	respectively. This is a slightly different construction compared to
	the one in Example \ref{ex:regldpc} and the resulting codes are
	better suited for the use of a \gls{wd}, see \cite[Design Rule
	1]{Iyengar2012}. Assume that transmission takes place using PM-QPSK
	in the linear regime and a \gls{wd} with $W = 10$ and $\maxIter = 7$
	is used. In Fig.~\ref{fig:ProtographExample} (a), we show the
	predicted \gls{ber} obtained via the \gls{pexit} analysis (solid
	lines) together with the actual performance of randomly generated
	codes (dashed lines) for two different lifting factors $\LF = 2000$
	(crosses) and $\LF = 10000$ (dots) for both the terminated (blue)
	and tailbiting (red) cases. Due to graph cycles, there is a mismatch
	between the actual performance and the \gls{de} prediction, in
	particular for the smaller lifting factor. However, the \gls{pexit}
	analysis accurately predicts the SNR region where the finite-length
	\gls{ber} curves ``bend'' into their characteristic waterfall
	behavior. \demo
\end{example}

In Fig.~\ref{fig:ProtographExample}~(a), we also indicate the two
points where the \gls{pexit} performance curves cross a \gls{ber} of
$10^{-5}$. We refer to the SNR value of such a point as the decoding
threshold $\rho^*$ for a target \gls{ber} $= 10^{-5}$ and a given
finite number of decoding iterations. The thresholds can be
numerically computed using a bisection search over a given SNR range.
The thresholds are given by $\rho^* \approx 0.82$ dB and $\rho^*
\approx 1.19$ dB for the terminated and tailbiting codes,
respectively.  \RevCR{(Alternatively, thresholds could also be given
in terms of the pre-FEC BER.)} The better decoding thresholds and
finite-length performance of the terminated code can be explained by
inspecting the structure of the base matrix
Fig.~\ref{fig:parity_check} (a). One may verify that the \gls{cn}
degrees corresponding to the first and last couple of rows are lower
than the CN degrees corresponding to the rows in between (see also
Fig.~\ref{fig:protograph_lifting} (a)). The lower degree CNs lead to a
locally better decoding capability, which is visualized by the colored
scale (green indicates a better correction capability), at the expense
of a rate loss. This termination boundary effect initiates the
wave-like decoding behavior that is characteristic for terminated
\gls{scldpc} codes \cite{Kudekar2011}. On the other hand, for the
tailbiting case, all CNs have the same degree $J$, hence no rate loss
is incurred and all positions are protected equally. However, this
also prevents the initiation of a decoding wave. 



\section{SC-GLDPC Codes with BCH Component Codes}
\label{sec:scgldpc}

\subsection{Code Construction}
\label{sec:scgldpc_construction}


We consider the $(\Cbch, \CnNumSp, \TL, w)$ \gls{scgldpc} code
ensemble proposed in \cite{Jian2012}, where $\Cbch$ is a binary linear
code of length $\nbch$ and dimension $\kbch$ that can correct all
error patterns of weight at most $t$,  $\CnNumSp$ is the number of
\CNs per spatial position, $\TL$ is the spatial length, and $w$ is the
coupling size. In the following, we assume that $\Cbch$ is a shortened
primitive \gls{bch} code with parameters $(\nbch, \kbch) = (2^{\nu}-1
- s,2^{\nu}-\nu t - 1 - s)$, where $\nu$ is the Galois field extension
degree and $s$ is the number of shortened information bits. The code
$\Cbch$ defines the constraints that have to be satisfied by all \CNs
in the Tanner graph representing the \gls{scgldpc} code.


For completeness, we review the construction of the terminated case
described in \cite[Def.~2]{Jian2012} and explain the necessary
modifications for the tailbiting case. Assume that $\CnNumSp$ \CNs
with degree $\nbch$ are placed at each of the spatial positions $1$ to
$\TL+w-1$ and $\CnNumSp \nbch /2$ \VNs of degree $2$ are placed at
each of the spatial positions $1$ to $\TL$. Additionally, $\CnNumSp
\nbch /2$ \VNs initialized to a known value are placed at positions $j
< 1$ and $j > \TL$ to terminate the code. The connections between \CNs
and \VNs are as follows. The $\CnNumSp \nbch$ VN and CN sockets at
each position are partitioned into $w$ groups of equal size $\CnNumSp
\nbch /w $ via a uniform random permutation.  The $i$-th group at the
$j$-th VN position and the $i$-th group at the $j$-th CN position are
denoted by $\mathcal{S}^{\text{(v)}}_{j,i}$ and
$\mathcal{S}^{\text{(c)}}_{j,i}$, respectively, where $i \in \{0, 1,
\dots, w-1\}$. The Tanner graph of one particular code in the ensemble
is constructed by using a uniform random permutation to connect
$\mathcal{S}^{\text{(v)}}_{j,i}$ to
$\mathcal{S}^{\text{(c)}}_{j+i,w-i-1}$ and mapping the $\CnNumSp \nbch
/w $ edges between the two groups. For the tailbiting case, the
position index $j+i$ is interpreted modulo $\TL$ and no known \VNs are
present. 



\begin{figure}[tb]
	\centering
		\includegraphics{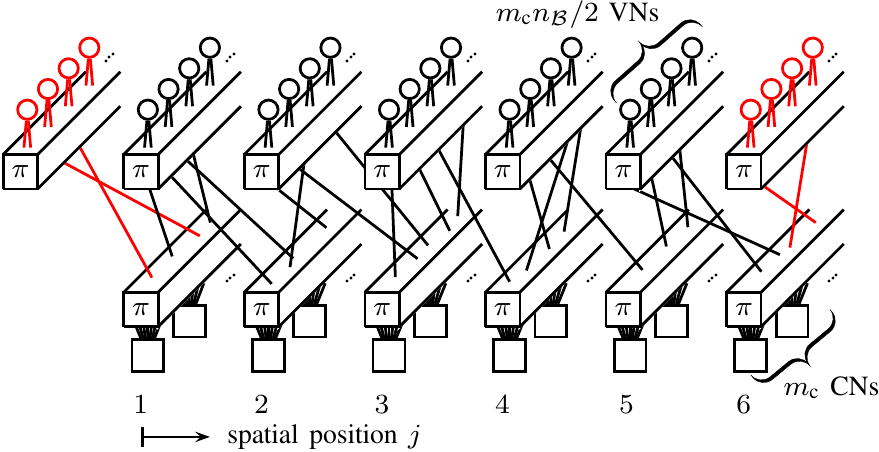}

	\caption{Tanner graph for a code in the
	terminated \gls{scgldpc} ensemble with $\TL = 5$ and $w = 2$. Known
	VNs are shown in red. }
	\label{fig:sc_ensemble}
\end{figure}

\begin{example}
	\label{ex:gldpc_construction}
	Consider the case where $\TL = 5$ and $w = 2$. The Tanner graph of a
	code in the terminated ensemble is shown in
	Fig.~\ref{fig:sc_ensemble}. The blocks $\pi$ spread out the edges
	from the \VNs and \CNs according to the random permutations in the
	construction. A code from the tailbiting ensemble is obtained by
	removing the known \VNs and the \CNs at position 6, and connecting
	the lose edges to the \CNs at position 1.  \demo
\end{example}


The design rate for the terminated ensemble is lower bounded by
\cite[eq.~(2.2)]{Yung-Yih2013}
\begin{align}
	\Rgldpc(\TL) \geq \Rgldpc - (1-\Rgldpc) \frac{w-1}{\TL},
	\label{eq:gldpc_rate}
\end{align}
where $\Rgldpc = 2 \kbch/\nbch - 1$ is the design rate for the
tailbiting ensemble. An exact expression for the design rate can be
obtained by explicitly considering the possibility that certain CNs
are connected exclusively to known VNs, similar to \cite[Lemma
3]{Kudekar2011}. However, for the high CN degrees and small coupling
factors considered in this paper, one can safely ignore this
possibility and we henceforth interpret \eqref{eq:gldpc_rate} as an
equality.

\begin{example}
	\label{ex:gldpc_rate}
	Let $\Cbch$ be a shortened \gls{bch} code with $\nu = 7$, $t = 3$,
	and $s = 43$, i.e., $\Cbch$ has rate $0.75$.  For the terminated and
	tailbiting ensembles in Example \ref{ex:gldpc_construction}, the
	design rates are given by $\Rgldpc(\TL) = 0.4$ and $\Rgldpc = 0.5$,
	respectively.  \demo
\end{example}

Similar to the parity-check matrix of an \gls{ldpc} code, a
\gls{gldpc} code can be specified by an incidence matrix
\cite[p.~220]{Ryan2009}. The dimensions of the incidence matrix are
$\CnNumSp (\TL+w-1) \times \TL \CnNumSp \nbch / 2$ and $\CnNumSp \TL
\times \TL \CnNumSp \nbch / 2$ for the terminated and tailbiting
ensembles, respectively. 

\begin{example}
	Consider the case where $w = 2$ and $\TL = \infty$. Let $\nbch$ be
	even and $\CnNumSp = \nbch/2$. If the edge permutations are such
	that the semi-infinite incidence matrix is the one shown in
	\cite[p.~54]{Smith2011}, the code corresponds to a staircase code.
	In other words, staircase codes are contained in the terminated
	ensemble for a certain choice of parameters $w$, $\TL$, and
	$\CnNumSp$.  \demo
\end{example}

\subsection{Hard-Decision Decoding and Density Evolution}
\label{sec:scgldpc_decoding}

\newcommand{\PE}[2]{\ensuremath{q_{#1}^{(#2)}}}
\newcommand{\PDE}[2]{\ensuremath{p_{\text{e},#1}^{(#2)}}}
\newcommand{\PA}[2]{\ensuremath{a_{#1}^{(#2)}}}
\newcommand{\PB}[2]{\ensuremath{b_{#1}^{(#2)}}}
\newcommand{\APE}[2]{\ensuremath{\bar{p}_{#1}^{(#2)}}}
\newcommand{\APA}[2]{\ensuremath{\bar{a}_{#1}^{(#2)}}}
\newcommand{\APB}[2]{\ensuremath{\bar{b}_{#1}^{(#2)}}}
\newcommand{\FOO}[1]{\ensuremath{f_{\nbch}^{1 \to 1}\left(#1\right)}}
\newcommand{\FZO}[1]{\ensuremath{f_{\nbch}^{0 \to 1}\left(#1\right)}}
\newcommand{\AvgOver}[1]{\ensuremath{\frac{1}{w} \sum_{#1 = 0}^{w-1}}}
\newcommand{\COP}{\ensuremath{p}} 

We use the iterative \gls{hdd} algorithm based on extrinsic message
passing of binary messages that is proposed in
\cite[Sec.~II-A]{Jian2012} (see also \cite[Sec.~II-C]{Jian2014}).
Assume transmission over a \gls{bsc} with crossover probability
$\COP$.  All outgoing \gls{vn} messages are initialized to the channel
observation. For each \gls{cn}, the incoming messages from the \VNs
are collected in a candidate decoding vector, which is then decoded
using \gls{bdd}. The outgoing \gls{cn} messages are computed based on
the Hamming distance between the candidate vector and the decoded
vector, cf.~\cite[Algorithm 1]{Jian2014}. In the next iteration, the
outgoing \gls{vn} message on a particular edge corresponds to the
incoming message on the other edge of that \gls{vn}.  Decoding
continues for $\maxIter$ iterations. The final decision for each
\gls{vn} is made based on the channel observation and the two incoming
messages. If the two messages agree, the bit is set to the message
value. If the messages disagree, the bit is set to the binary
complement of the channel observation.  As pointed out in
\cite{Jian2012}, extrinsic message passing is different
compared to the conventional approach of decoding product-like codes
(referred to as intrinsic message passing in \cite{Jian2012}) and can
be rigorously analyzed via \gls{de} even in the event of miscorrection
\cite{Jian2012}, i.e., when undetected errors remain after
\gls{bdd}.

We briefly summarize the \gls{de} analysis presented in
\cite{Jian2012}. Assume that the all-zero codeword is transmitted and
let $\PE{j}{l}$ be the average probability that a message emitted by a
\gls{vn} at position $j$ is in error (i.e., the message is ``1'')
after the $l$th iteration. The DE recursion is given by
\cite[eq.~(5)]{Jian2012}
\begin{align}
	\PE{j}{l} = \AvgOver{k} f_{\nbch} \left( \AvgOver{k'} \PE{j-k'+k}{l-1} ;
	\COP
	\right), 
	\label{eq:gldpc_de_bsc}
\end{align}
with \cite[eq.~(2)]{Jian2012}
\begin{align}
	\begin{split}
	f_{\nbch}(x;\COP) &\define \sum_{i=0}^{\nbch-1} \binom{\nbch-1}{i} x^i
	(1-x)^{\nbch-1-i} \\
	&\phantom{\define} \quad \cdot
	\left(
		\COP P_{\nbch}(i) + (1 - \COP) Q_{\nbch}(i)
	\right), 
\end{split}
	\label{eq:gldpc_de_bsc_f}
\end{align}
where $P_{\nbch}(i)$ and $Q_{\nbch}(i)$ are defined in
\cite[eq.~(3)]{Jian2012} and \cite[eq.~(4)]{Jian2012}.
The initial conditions are $\PE{j}{0} = \COP$ for $j \in \{1,\ldots,
\TL\}$ and $\PE{j}{l} = 0$ for $j \notin \{1, \ldots, \TL\}$.
For tailbiting ensembles, the subscript $j-k'+k$ in
\eqref{eq:gldpc_de_bsc} is calculated modulo $\TL$.

The analysis in \cite{Jian2012} is presented for unshortened \gls{bch}
codes, \IE $s=0$. However, \CNs connected to known variable nodes are
treated as shortened component codes by adjusting the effective error
probability of the incoming messages through the boundary condition
$\PE{j}{l} = 0$ for $j \notin \{1,\ldots, \TL\}$. If the component codes are
shortened \gls{bch} codes, one can therefore apply the same analysis
as before, where $\nbch$ now denotes the length of the unshortened
code and the function $f_{\nbch}(x; p)$ is replaced by $f_{\nbch}(x
(\nbch - s) / \nbch; p)$. 

The \gls{ber} for the \VNs at position $j$ after the $l$th iteration was not
derived in \cite{Jian2012}, but can be easily found as follows. First,
we rewrite \eqref{eq:gldpc_de_bsc_f} in the form
\begin{align}
	f_{\nbch}(x;\COP)	&= \COP \FOO{x} + (1-\COP) \FZO{x}, 
\end{align}
where $\FOO{x}$ and $\FZO{x}$ are implicitly defined via
\eqref{eq:gldpc_de_bsc_f}.
We introduce the two variables 
\begin{align}
	\PA{j}{l} \define \FOO{\AvgOver{k'} \PE{j-k'}{l}} 
	, \quad
	\PB{j}{l} \define \FZO{\AvgOver{k'} \PE{j-k'}{l}}
\end{align}
and their averages
\begin{align}
	\APA{j}{l} \define \frac{1}{w} \sum_{k=0}^{w-1} \PA{j+k}{l}  
	\quad \text{and} \quad
	\APB{j}{l} \define \frac{1}{w} \sum_{k=0}^{w-1} \PB{j+k}{l} . 
\end{align}
With these definitions, the recursion \eqref{eq:gldpc_de_bsc} becomes
\begin{align}
	\PE{j}{l} =& \ \COP \APA{j}{l-1}
	+(1-\COP) \APB{j}{l-1}
\end{align}
and the decoding error probability after the $l$th iteration is 
\begin{align}
	\PDE{j}{l} =& \ \COP \left(\APA{j}{l-1}\right)^2
	+(1-\COP) \left(1- \left(1-\APB{j}{l-1}\right)^2\right).
\end{align}
The final \gls{ber} after $\lmax$ steps of iterative \gls{hdd} is computed
as $p_e = \frac{1}{\TL} \sum_{j=1}^{\TL} \PDE{j}{\lmax}$.


%


Since we intend to use the \gls{de} analysis in an optimization
routine, we approximate the two functions $\FOO{x}$ and $\FZO{x}$ with
their high-rate scaling limit versions (i.e., for $\nbch \to \infty$)
which are easier to compute and given by \cite{Jian2012} 
\begin{align}
	\label{eq:gldpc_de_bsc_f11_approx}
	\FOO{x} \approx \phi\left( \nbch x ; t - 1\right)
\end{align}
and
\begin{align}
	\label{eq:gldpc_de_bsc_f01_approx}
	\FZO{x} \approx \frac{1}{\nbch (t-1)!} \phi\left( \nbch x ;
	t\right),
\end{align}
where $\phi\left( \lambda ; t\right) = 1 - \sum_{i=0}^{t}
\frac{\lambda^i}{i!} e^{-\lambda}$. 

It is straightforward to modify the decoding algorithm and the
\gls{de} analysis if a similar \gls{wd} as for the protograph-based
\gls{scldpc} codes is used and hence we omit the details. 

\begin{example}
	\label{ex:gldpc}
	Consider the case where $\TL = 20$ and $w = 2$. Let $\Cbch$ be the
	same \gls{bch} code as in Example \ref{ex:gldpc_construction}. The
	design rates are $\Rgldpc(\TL) = 0.475 $ and $\Rgldpc = 0.5$,
	respectively. Assume transmission using PM-QPSK in the linear regime
	and a \gls{wd} with $W = 5$ and $\maxIter = 10$. In
	Fig.~\ref{fig:ProtographExample} (b), we show the predicted
	\gls{ber} obtained via \gls{de} (solid lines) together with the
	actual performance of randomly generated codes (dashed lines) for
	$\CnNumSp = 200$ (crosses) and $\CnNumSp = 1500$ (dots) for both the
	terminated (blue) and tailbiting (red) cases. The decoding
	thresholds at a \gls{ber} of $10^{-5}$ are $\rho^* \approx 3.71$
	dB and $\rho^* \approx 3.94$ dB, respectively.  \demo
\end{example}

\section{Bit Mapper Optimization}
\label{sec:optimization}

The different modulation bits of a nonbinary signal constellation have
different protection levels, which can be taken advantage of by
optimizing the bit mapper. This concept is easiest to understand for
\gls{hdd}, which we describe first. 

For the ``hard'' demodulator, the entire block diagram shown in
Fig.~\ref{fig:BlockDiagramConf} can be replaced by $m$ parallel
\glspl{bsc} with different crossover probabilities $p_i$, $1 \leq i
\leq m$, which depend on the signal constellation, binary labeling,
and \gls{snr} $\rho$. Each VN corresponds to a coded bit, and for the
\gls{scgldpc} codes there are $\CnNumSp \nbch / 2 $ VNs at each
spatial position (see Section \ref{sec:scgldpc_construction}). The
baseline bit mapper (see Section \ref{sec:system_model}) allocates the
same number of coded bits from each spatial position to the different
modulation bits (i.e., the $m$ parallel \glspl{bsc}). In this case,
the crossover probability for the VNs at an arbitrary spatial position
is simply the average $\bar{p} = \frac{1}{m} \sum_{i=1}^{m} p_i $.
More generally, the bit mapper is modeled by specifying the
assignment of VNs to the modulation bits via a matrix $\mathbf{A} =
[a_{i,j}] \in \mathbb{R}^{m \times \TL}$, where $a_{i,j}$, $0 \leq
a_{i,j} \leq 1$ $\forall i, j$, denotes the proportional allocation of
the coded bits corresponding to the VNs at spatial position $j$
allocated to the $i$th modulation bit, and $\sum_{i=1}^{m} a_{i,j} =
1$, for all $j$.  The effective crossover probability for the \VNs at
spatial position $j$ is therefore a weighted average of the \gls{bsc}
crossover probabilities according to $\eps_j = \sum_{i=1}^{m} a_{i,j}
p_i$. To account for different crossover probabilities at the spatial
positions in the \gls{de} analysis, we can simply replace $\COP$ in
\eqref{eq:gldpc_de_bsc} by $\eps_j$.




For the protograph-based \gls{scldpc} codes with \gls{sdd}, one can
make similar considerations. Each VN in the protograph represents
$\LF$ \VNs in the lifted Tanner graph, \IE $\LF$ coded bits. If we
assume for example that a given protograph is lifted with a lifting
factor $\LF$ which is divisible by $m$, the baseline bit mapper
allocates $\LF/m$ coded bits for each protograph VN to each modulation
bit. The bit mapper is modeled via a matrix $\mathbf{A} = [a_{i,j}]
\in \mathbb{R}^{m \times \nldpc'}$, where $a_{i,j}$ now denotes the
proportional allocation of the coded bits corresponding to the $j$th
column in the base matrix allocated to the $i$th modulation bit. The
matrix $\mathbf{A}$ is then used in the modified \gls{pexit} analysis
to predict the iterative performance behavior under \gls{sdd}
\cite[Algorithm 1]{Hager2014a}. 

We optimize $\mathbf{A}$ based on the decoding threshold with the help
of differential evolution \cite{Storn1997}. For more details about the
optimization procedure, we refer the reader to \cite{Hager2014a},
where we also discuss several techniques to reduce the optimization
complexity for \gls{sc} codes. Once an optimized bit mapping matrix
$\vect{A}^*$ is found, the finite-length bit mapper is obtained via
the rounded matrix $(\CnNumSp \nbch / 2) \vect{A}^*$ for the
\gls{scgldpc} codes and via $\LF \vect{A}^*$ for the \gls{scldpc}
codes, from which the index assignment of coded bits to modulation
bits is determined.

\section{Results and Discussion}
\label{sec:results}

Since this paper does not deal with code design, we rely on code
parameters that have been proposed elsewhere in the literature in
order to illustrate the bit mapper optimization technique. For the
numerical results, we consider protograph-based \gls{scldpc} codes
with $\proto_0 = (1, 2, 1, 2)$ and $\proto_1 = (3, 2, 3, 2)$, where
$J' = 1$, $K' = 4$, and $m_s = 1$ \cite{Schmalen2012}. The design rate
of the tailbiting case is $\Rldpc = 0.75$ ($\text{OH} = 33$\%). For
the \gls{scgldpc} codes, we restrict ourselves to $w = 2$ and use the
BCH code parameters in \cite[Table I]{Zhang2014} which are optimized
for staircase codes. In particular, we consider $\nu = 9$, $t = 4$,
and $s = 223$, which again leads to $\Rgldpc = 0.75$.  The staircase
code for these parameters is estimated to perform approximately $1.38$
dB away from the \gls{bsc} capacity (at a \gls{ber} of $10^{-15}$)
\cite[Table I]{Zhang2014}. We also consider an example with higher
rate and performance closer to capacity. In particular, we consider
$\nu = 10$, $t = 4$, $s = 143$ where $\Rgldpc = 0.91$ ($\text{OH} =
10$\%). The gap to capacity of the staircase code for these parameters
is estimated to be $0.59$ dB \cite[Table I]{Zhang2014}.

The bit mapper optimization is performed for the terminated and
tailbiting cases of the three code examples for different spatial
lengths $\TL \in \{12, 21, 30, 39, 48, 57, 66, 75, 84, 300\}$.  We
consider Gray-labeled PM-$16$-QAM, PM-$64$-QAM, and PM-$256$-QAM. In
all scenarios, a \gls{wd} is employed with a window size of $\WS = 5$
and $\maxIter = 10$ iterations per window. The target \gls{ber} for
the optimization is set to $10^{-5}$. Setting a lower target \gls{ber}
(e.g., $10^{-15}$) has virtually no influence on the optimization
outcome due to the steepness of the predicted \gls{de} performance
curves (see Fig.~\ref{fig:ProtographExample}). This assumes that there
are no error floors due to harmful graph structures, which cannot be
modeled using \gls{de} and are not considered in this paper. An
analysis of the error floor for the considered \gls{scgldpc} codes is
an interesting topic for future work and beyond the scope of this
paper. 

\subsection{Structure of the Optimized Bit Mapper for Tailbiting
Codes}

\begin{figure}[t]
	\centering
		\subfloat[SC-LDPC, $\Rldpc=0.75$,
		PM-64-QAM]{\includegraphics[width=1.0\columnwidth]{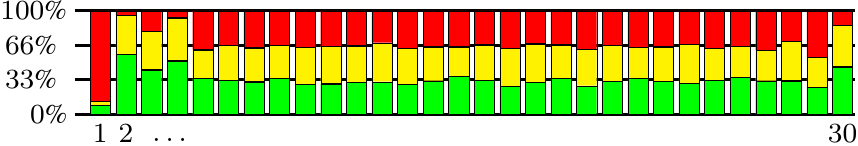}}

		\subfloat[SC-GLDPC, $\Rgldpc=0.75$,
		PM-16-QAM]{\includegraphics{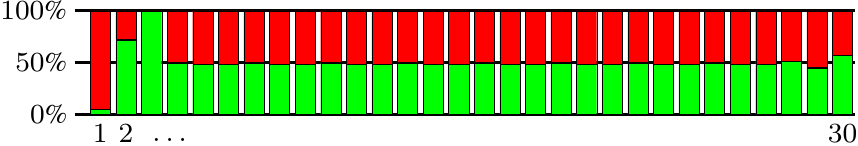}}

	\caption{Optimized allocation to the modulation bits with the best
	(green), worst (red), and intermediate (yellow, for PM-64-QAM)
	protection level for each spatial position of the tailbiting code in
	two scenarios.}

	\label{fig:BitMapper}
\end{figure}

For the tailbiting codes, the optimized bit mappers have an
interesting structure, which is illustrated in
Fig.~\ref{fig:BitMapper} for two scenarios: (a) the \gls{scldpc} code
with SDD, PM-64-QAM, and $\TL = 30$; (b) the \gls{scgldpc} code with
HDD, $\Rgldpc = 0.75$, PM-16-QAM, and $\TL = 30$. For PM-64-QAM and
PM-16-QAM, the modulation bits have three and two different protection
levels, respectively. Due to the tailbiting code structure, the
bit allocation is invariant to a circular shift, assuming that the
scheduling of the \gls{wd} is modified according to the same shift.
For the allocation shown in Fig.~\ref{fig:BitMapper}, it is assumed
that the first decoding window begins at the first spatial position.
The optimized bit mapper in both scenarios deviates significantly from
the baseline mapper in the first few spatial positions. For the
\gls{scldpc} code, the coded bits corresponding to the second, third,
and fourth spatial position are proportionally more allocated to the
best (green) and intermediate (yellow) protection level of PM-64-QAM.
Similarly, for the \gls{scgldpc} code, the coded bits corresponding to
the second and third spatial position are proportionally more
allocated to the best modulation bit of PM-16-QAM. In both cases, the
optimized allocation leads to a locally improved decoding convergence
and initiates a wave-like decoding behavior comparable to that of
terminated codes, i.e., the unequal error protection of the signal
constellation is exploited to create an artificial termination
boundary.   

The performance gain due to the optimized bit mapper (which is
quantified in the next section) comes at the expense of some increase
in system complexity. In particular, one has to account for additional
buffering because a symbol cannot be transmitted until all its $m$
bits are encoded. For simplicity, let us assume a model where the FEC
encoder outputs coded bits in blocks of $M K'$ or $\CnNumSp \nbch /2$
bits, i.e., the number of bits per spatial position, and symbols are
immediately modulated as soon as all $m$ modulation bits are
available. Then, no buffering is required for the sequential baseline
mapper. On the other hand, the ``worst-case'' bit mapper allocates
100\% of the coded bits in the first $\TL/m$ spatial positions to the
first modulation bit, 100\% in the next $\TL/m$ positions to the
second bit, and so on (i.e., $a_{i,j} = 1$ for $(i-1)T/m+1 \leq j \leq
i\TL/m$). Consequently, no bits are allocated to the last modulation
bit until spatial position $(m-1)\TL/m+1$ and buffering of all coded
bits up to position $(m-1)\TL/m$ is required. In all considered
scenarios, however, the required additional buffer size (in terms of
the number of spatial positions) due to the optimized bit mappers did
not exceed 2. 


\subsection{Optimization Gain}

\begin{figure*}[t]
	\centering
		\subfloat[SC-LDPC, $\Rldpc = 0.75$]{\includegraphics{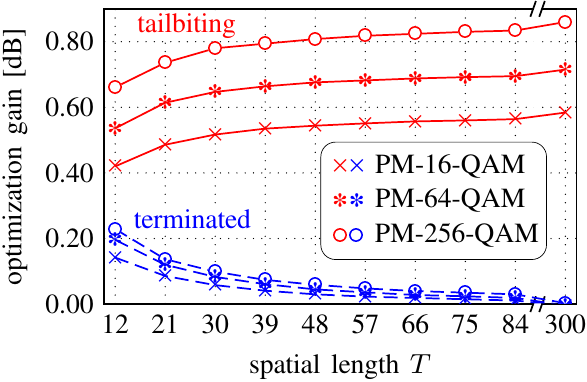}}
		\quad
		\subfloat[SC-GLDPC, $\Rgldpc = 0.75$]{\includegraphics{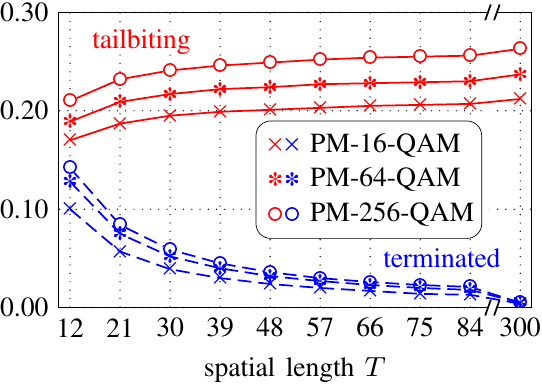}}
		\quad
		\subfloat[SC-GLDPC, $\Rgldpc = 0.91$]{\includegraphics{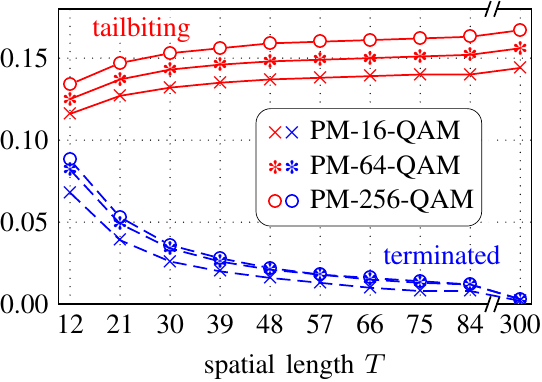}}

	\caption{Optimization gain as a function of the spatial length $\TL$.}

	\label{fig:gain}
\end{figure*}

\begin{figure*}[t]
	\centering
		\subfloat[SC-LDPC, $\Rldpc = 0.75$]{\includegraphics{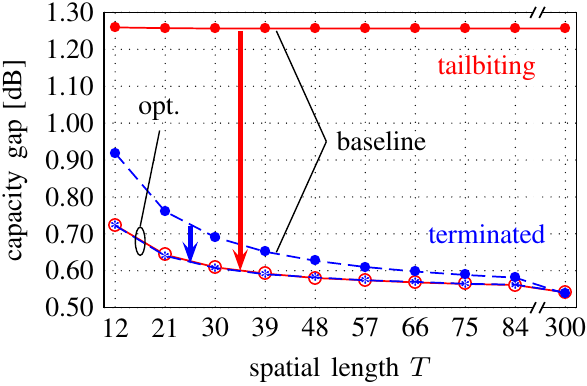}}
		\quad
		\subfloat[SC-GLDPC, $\Rgldpc = 0.75$]{\includegraphics{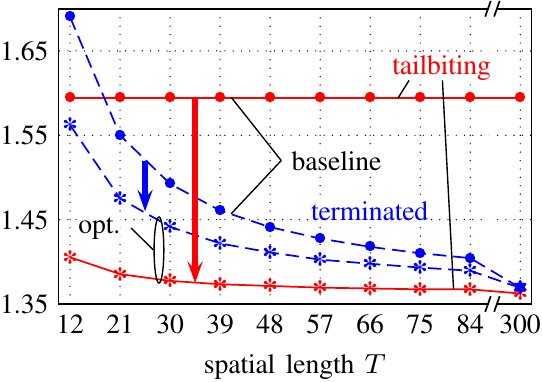}}
		\quad
		\subfloat[SC-GLDPC, $\Rgldpc = 0.91$]{\includegraphics{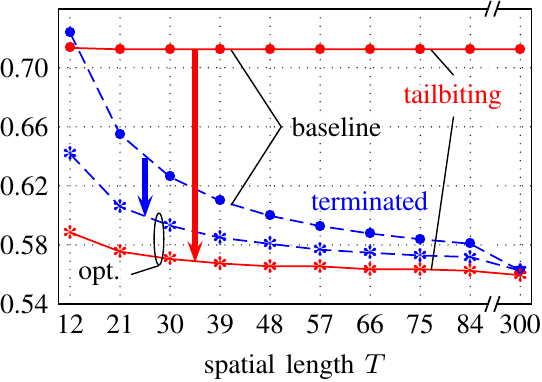}}

	\caption{Capacity gap as a function of the spatial length $\TL$ for
	PM-64-QAM.}

	\label{fig:gap}
\end{figure*}

In Fig.~\ref{fig:gain}, we show the optimization gain (in dB) as a
function of the spatial length $\TL$ for all considered scenarios. The
optimization gain is defined as the difference between the decoding
threshold using the baseline bit mapper and the decoding threshold
using the optimized bit mapper. The gain quantifies the performance
improvement one can expect by employing the optimized bit mappers
assuming long codes. 

Regardless of the signal constellation or code class, the optimization
gain decreases with $\TL$ for the terminated codes and increases for
the tailbiting codes. This behavior can be explained as follows. The
optimization gain for the tailbiting codes comes from allocating more
coded bits in the beginning of the spatial chain to good modulation
bits in order to initiate a decoding wave. This, however, reduces the
effective capacity for the bits in the middle part of the spatial
chain. As $\TL$ increases, this reduction becomes negligible and the
optimization gain tends to a constant value. For terminated codes, a
decoding wave is initiated by default and the optimized bit mapper
increases the effective capacity for the bits in the middle part by
allocating bits in the beginning and end of the chain proportionally
more to modulation bits with lower protection levels. Again, as $\TL$
increases, this effect becomes negligible and the gain approaches
zero. As a result, while the tailbiting codes significantly benefit
from the optimization, the gain for the considered terminated codes is
limited, i.e., for $\TL \geq 30$ the gain is $< 0.1$ dB in all cases. 

It can also be observed that the optimization gain generally depends
on the signal constellation. The gain increases with the modulation
order $M$ due to the increased number of protection levels and
stronger unequal error protection. This gain increase can also be
observed when optimizing bit mappers for irregular \gls{ldpc} codes,
see, \EG \cite{Cheng2012}. It is also important to stress that
the optimization relies on the availability of a signal constellation
with different protection levels in order to provide a performance
gain. In particular, the techniques do not apply to PM-BPSK or
(Gray-labeled) PM-QPSK. 


\subsection{Gap to Capacity}

In order to gain some insight into the performance of the terminated
and tailbiting codes relative to each other, the capacity gap (in dB)
as a function of the spatial length $\TL$ is shown in
Fig.~\ref{fig:gap} for PM-64-QAM. For \gls{sdd} of the \gls{scldpc}
codes, the BICM capacity \cite{Caire1998} is taken as a benchmark.
For HDD of the \gls{gldpc} codes, the capacity of the \gls{bsc}
with averaged crossover probability is taken as a benchmark, similar
to \cite{Smith2012}.  Alternatively, one may use the capacity of the
sum of the $m$ parallel \glspl{bsc} as a benchmark, which is larger.
The gains discussed in the previous subsection are indicated in
Fig.~\ref{fig:gap} with arrows.

The decoding thresholds for the baseline systems are approximately
independent of $\TL$. Therefore, the capacity gap for the tailbiting
codes remains constant in all cases, while the capacity gap for the
terminated codes decreases due to the vanishing rate loss. For the
baseline systems, the performance difference between terminated and
tailbiting codes is most significant for the \gls{scldpc} codes (up to
$0.75$ dB), while for the \gls{scgldpc} codes the difference is lower
(up to $0.25$ dB for $\Rgldpc = 0.75$ and up to $0.19$ dB for $\Rgldpc
= 0.91$). In all cases, the capacity gap is reduced by employing the
optimized bit mappers. If we compare the optimized systems, it can be
seen that the gap for the \gls{scldpc} codes is virtually identical
for terminated and tailbiting cases. For the \gls{scgldpc} codes, the
tailbiting codes perform closer to capacity, albeit the difference to
the terminated codes for $\TL \geq 30$ is small. For very long spatial
lengths (i.e., $\TL = 300$), the capacity gap virtually overlaps also
for the \gls{scgldpc} codes. 


\subsection{Simulation Results}
\label{sec:results_scldpc}

The results presented in the previous two subsections are based on
decoding thresholds, i.e., assume an infinite code length. The
deviation of the \gls{de} analysis from the finite-length performance
is determined by the lifting factor $\LF$ and the number of CNs per
position $\CnNumSp$, see Fig.~\ref{fig:ProtographExample}.



\begin{figure}
	\centering
		\includegraphics{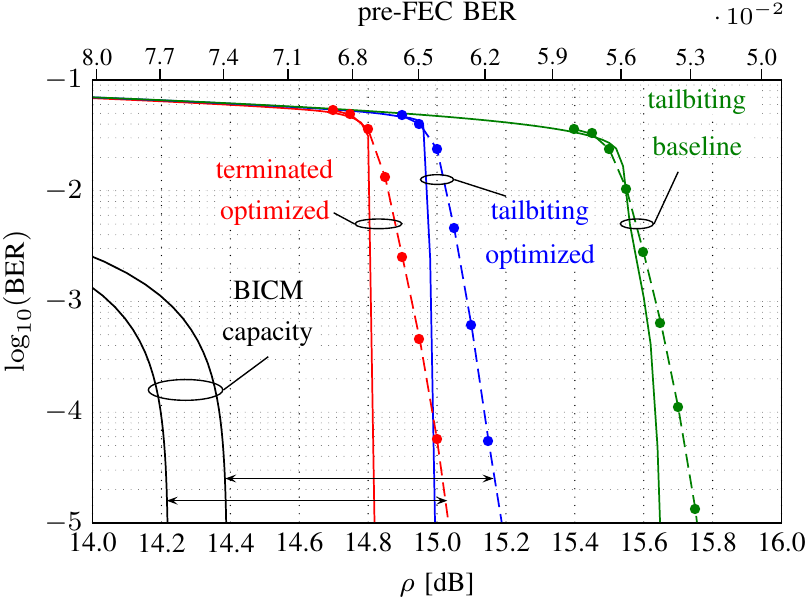}

	\caption{Simulation results for the SC-LDPC code with PM-64-QAM over
	the AWGN channel. The codes have length $360\,000$.}
	\label{fig:ComparisonAWGN}
\end{figure}

As an example, consider the \gls{scldpc} code with $T = 30$ and
$\LF=3000$ leading to a code length of $\nldpc = 360\,000$. The rates
are $\Rldpc(30) \approx 0.742$ and $\Rldpc = 0.75$, respectively.  In
Fig.~\ref{fig:ComparisonAWGN}, we show simulation results (dashed
lines with dots) and the analytical P-EXIT prediction (solid lines)
for the AWGN channel, i.e., a linear transmission scenario, assuming
PM-64-QAM. As predicted by the optimization gain in
Fig.~\ref{fig:gain} (a), the tailbiting code performs significantly
better with an optimized bit mapper and a gain of $\approx 0.55$ dB is
achieved at a \gls{ber} of $10^{-5}$. The terminated code performs
better for the same \gls{snr}, but entails a smaller spectral
efficiency due to the rate loss. The gap to the \gls{ber}-constrained
BICM capacity \cite[p.~17]{Ryan2009} of the two optimized systems, as
indicated by the arrows and predicted from Fig.~\ref{fig:gap} (a), is
approximately the same (as is the gap to the AWGN channel capacity,
not shown). 

\begin{figure}[t]
	\centering
		\includegraphics{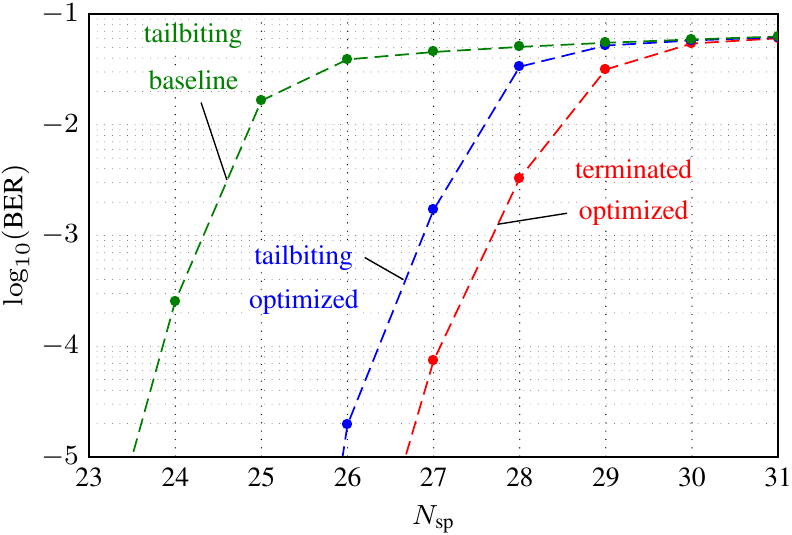}

	\caption{Simulation results for the SC-LDPC with
	PM-64-QAM
	over a dispersion uncompensated transmission link.}
	\label{fig:ComparisonNLSE}
\end{figure}

Lastly, we also present simulation results for a nonlinear
transmission scenario. We set $\alpha = 0.25$ dB/km, $\beta_2 = -
21.668$ ps${ }^2$/km, $\gamma = 1.4$ W${ }^{-1}$ km${ }^{-1}$, $\nu_s
= 1.934 \times 10^{14}$ Hz, $\SEF = 1.622$, $R_s = 40$ GBaud, and
$\Lsp = 70$ km. A root-raised cosine pulse $p(t)$ with a roll-off
factor of $0.25$ is used. We employ the symmetric \acl{ssfm} with two
samples per symbol and a fixed step size
\cite[Sec.~2.4.1]{Agrawal2006}. The input power per polarization is
set to $P = -2.5$ dBm. In the simulation model, the polarization state
is assumed to be known and perfect timing and carrier synchronization
is assumed. In Fig.~\ref{fig:ComparisonNLSE}, the simulated \gls{ber}
of the \gls{pm} transmission systems is plotted as a function of the
number of fiber spans $\Nsp$. For the tailbiting code, the $0.55$ dB
gain obtained by using the optimized bit mapper translates into an
increase of the transmission reach by roughly 3 additional spans or
approximately $13$\%. This gain is obtained at almost no
increased system complexity cost, i.e., by simply replacing the
baseline bit mapper with an optimized one. This reach extension can
also be approximately calculated using the analytical expression for
the SNR $\rho$ as a function of the number of spans presented in
\cite{Beygi2012}. The terminated code enables a longer transmission
reach of approximately one span, at the expense of a $1.2$\% decrease
in spectral efficiency.  The performance of the terminated code with
the baseline bit mapper is very close to the performance of the
tailbiting code with the optimized bit mapper and is therefore not
shown in Figs.~\ref{fig:ComparisonAWGN} and \ref{fig:ComparisonNLSE}
for clarity. 

\section{Conclusions}
\label{sec:conclusion}

In this paper, we considered the optimized allocation of the coded
bits from the FEC encoder to the modulation bits for terminated and
tailbiting \gls{sc} \gls{fec} schemes, assuming both \gls{sdd} and
\gls{hdd}, as well as different signal constellations. Terminated
\gls{sc} codes generally benefit little from the optimization,
particularly for long spatial lengths. However, the performance of
tailbiting \gls{scldpc} codes can be significantly improved. With an
optimized bit allocation, the terminated and tailbiting codes are
competitive, in the sense that spectral efficiency can be traded for
transmission reach, at approximately the same gap to capacity.

\section{Acknowledgements}

The authors would like to acknowledge helpful discussions with Henry
D.~Pfister regarding the \gls{scgldpc} ensemble. 



\end{document}